\documentclass[11pt]{article}
\usepackage{amssymb, amsmath}
\begin{document}
\newcommand{\bm}[1]{\mbox{\boldmath $#1$}}

\makeatletter
%%%%%------- preprint style ----%%%%%%%%%%%%%%%%%%%%
\@addtoreset{equation}{section}
\def\theequation{\thesection.\arabic{equation}}
\def\@maketitle{\newpage
 \null
 {\normalsize \tt \begin{flushright} 
  \begin{tabular}[t]{l} \@date  
  \end{tabular}
 \end{flushright}}
 \begin{center} 
 \vskip 2em
 {\LARGE \@title \par} \vskip 1.5em {\large \lineskip .5em \begin{tabular}[t]{c}\@author 
 \end{tabular}\par} 
 \end{center}
 \par
 \vskip 1.5em} 
\makeatother
%%%%%%%%%%%%%%%%%%%%%%%%%
\topmargin=-1cm
\oddsidemargin=1.5cm
\evensidemargin=-.0cm
\textwidth=15.5cm
\textheight=22cm
%\renewcommand{\baselinestretch}{1.5}
%%%%%%%%%%%%%%%%%%%%%%%%
\setlength{\baselineskip}{16pt}
\title{A Bulk Localized State  and New Holographic 
Renormalization Group Flow in 3D Spin-3 Gravity
}
%}
\author{
Ryuichi~{\sc Nakayama}\thanks{nakayama@particle.sci.hokudai.ac.jp},  \ \  \  and \ \ \ Tomotaka~{\sc Suzuki}\thanks{t-suzuki@particle.sci.hokudai.ac.jp} 
       \\[1cm]
{\small
    Division of Physics, Graduate School of Science,} \\
{\small
           Hokkaido University, Sapporo 060-0810, Japan}
}
\date{
EPHOU-17-017  \\
December  2017 
}
% 
%\begin{titlepage}
% 
\maketitle

\begin{abstract} 
We construct a localized state of a scalar field in 3D spin-3 gravity. 
3D  spin-3 gravity is thought to be holographically dual to W$_3$ extended CFT on a boundary at infinity. 
It is known that while W$_3$ algebra is a non-linear algebra, in the limit of large central charge $c$ a linear finite-dimensional subalgebra generated by $W_n \, (n=0,\pm1,\pm2)$ and $L_n (n= 0,\pm1)$ is singled out. 
The localized state is constructed in terms of these generators. 
To write down an equation of motion for a scalar field which is satisfied by this localized state it is necessary to introduce new variables for an internal space $\alpha^{\pm}$, $\beta^{\pm}$, $\gamma$, in addition to ordinary coordinates $x^{\pm}$ and $y$. The higher-dimensional space, which combines the bulk spacetime  with  the `internal space', which is an analog of superspace in supersymmetric theory, is introduced. The `physical bulk  spacetime' is a 3D hypersurface with constant $\alpha^{\pm}$, $\beta^{\pm}$ and $\gamma$ embedded in this space. 
We  will work in Poincar\'e coordinates of AdS space and  consider W-quasi-primary operators $\Phi_{\bm{h}}(x^+)$ with a conformal weight $\bm{h}$ in the boundary and study two and three point functions of W-quasi-primary operators transformed  as $e^{ix^+L^h_{-1}} e^{\beta^+W^h_{-1}} \Phi_{\bm{h}}(0) e^{-\beta^+W^h_{-1}}e^{-ix^+L^h_{-1}}$. Here $L^h_n$ and $W^h_n$ are sl(3,R) generators in the hyperbolic basis for Poincar\'e coordinates. It is shown that in the $\beta^+ \rightarrow  \infty$ limit, the conformal weight changes to a new value $\bm{h}'=\bm{h}/2$. This may be regarded as a Renormalization Group (RG) flow. It is argued that this RG flow will be triggered by terms $\Delta S \propto \beta^+ W^h_{-1}+\beta^- \overline{W}^h_{-1}$ added to the action.

\end{abstract}
%\end{titlepage}
\newpage
\setlength{\baselineskip}{18pt}

%%%%%%%%%%%%%%%%%%%%%%%%%%%%%%%%%%%%%%%%%%%%%%%%%%%%%%%%%%%%%%%%%%%%%
%\newcommand{\bm}[1]{\mbox{\boldmath $#1$}}
%%%%%%%%%%%%%%%%%%%%%%%%%%%%%%%%%%%%%%%%%%%%%%%%%%%%%%%%%%%%%%%%%%%%
\section{Introduction}
\hspace{5mm}
Higher spin theory is a toy model of string theory. In \cite{GKP} 3d O(N) vector theory is shown to be dual to 4d Vasiliev theory\cite{Vasiliev}. In 3d it is possible to truncate higher spins to $s \leq N$ and the dynamics of higher spin theory can be described in terms of sl(N,R) $\oplus$ sl(N,R) Chern-Simons (CS) gauge theory.\cite{Campoleoni} In \cite{GG}\cite{GG2} minimal model of higher spin theory was proposed by using coset construction. Black hole solution was obtained in \cite{BH}.

We will focus on spin 3 gravity and W$_3$ algebra. Spin-3 gravity in three dimensions can be formulated as sl(3, R) $\oplus$ sl(3,R) Chern-Simons (CS) gauge theory.\cite{Deser}\cite{W}\footnote{These Lie algebras are related to generalizations of 3d diffeomorphism and local Lorentz transformations. They should {\it not} be confused with those associated with the chiral and anti-chiral W$_3$ algebras on the 2d boundary, which will be henceforth discussed in this paper.  } This is considered to be dual to W$_3$-extended CFT$_2$ on the boundary. \cite{Campoleoni} The W$_3$-extended conformal algebra is a non-linear algebra\cite{Zamolodchikov}\cite{BouwknegtSchoutens} defined by
\begin{eqnarray}
\ [ L_m, L_n] &=& (m-n) \, L_{m+n}+\frac{c}{12}m(m^2-1) \,  \delta_{m+n,0}, \nonumber \\
\ [L_m, W_n] &=& (2m-n) \, W_{m+n}, \nonumber \\
\ [ W_m, W_n] &=& -\frac{c}{36}m(m^2-1)(m^2-4) \, \delta_{m+n,0}\nonumber \\  &&-\frac{1}{3}(m-n)(2m^2+2n^2-mn-8) \, L_{m+n} \nonumber \\ &&-10\beta (m-n)\Lambda_{m+n}, \qquad (m,n \in {\bf Z}). \label{nonlinerW3}
\end{eqnarray}
 Here $\Lambda_m$ is a normal-ordered operator 
\begin{eqnarray}
\Lambda_m &=& \sum_{n=-\infty}^{\infty}(L_{m-n}, L_n)-\frac{3}{10}(m+3)(m+2) \, L_m, \nonumber \\
&=& \sum_{n \leq -2}L_n L_{m-n}+\sum_{n \geq -1}L_{m-n}L_n-\frac{3}{10}(m+3)(m+2) \, L_m,
\end{eqnarray}
and $\beta$ is a constant related to the central charge $c$:
\begin{equation}
\beta = \frac{16}{22+5c}. \label{betac}
\end{equation}
The normalization of $W_n$ is modified compared to that in \cite{BouwknegtSchoutens} by multiplying the right hand side of the equation for commutator $[W_m,W_n]$ by a factor $-10 $.\footnote{This makes the central charge term of this commutator negative for $c>0$. If we adopt a rule of hermitian conjugation, $(W_n)^{\dagger}=W_{-n}$,  this creates negative norm states.  Even if we define $(W_n)^{\dagger}=-W_{-n}$, there are negative norm states, since in the semi-classical limit  the algebra gives $[W_2,W_{-2}]=-16L_0$ and $[W_1,W_{-1}]=2L_0$.  See (\ref{W3wedgealgebra}). This will be important, when applying  the c-theorem\cite{c-th}.  In this paper this issue will not be studied.}  We will adopt the rule of hermitian conjugation $(W_n)^{\dagger}=W_{-n}$.\footnote{To switch to the other rule $(W_n)^{\dagger}=-W_{-n}$  we need to make a replacement $W_n \rightarrow i W_n$.This change can alternatively be performed by replacement of parameters $\alpha^{\pm}, \beta^{\pm}, \gamma \rightarrow i\alpha^{\pm}, i\beta^{\pm}, i\gamma$ in (\ref{Plocalized}) in sec. 4.}

In the semi-classical limit of the bulk theory $c \rightarrow \infty$, $\beta$ (\ref{betac}) is ${\cal O}(c^{-1})$, and by dropping the 3rd term in (\ref{nonlinerW3}), we can restrict discussion to the sub-algebra of the wedge modes, $L_n \ (n=0, \pm 1)$ and $W_m \ (m=0, \pm 1, \pm2)$. It is a linear sl(3,R) algebra and given by 
\begin{eqnarray}
\ [L_m,L_n] &=& (m-n) \, L_{m+n}, \qquad 
\ [L_m,W_n] = (2m-n) \, W_{m+n}, \nonumber \\
\ [W_m,W_n] &=& -\frac{1}{3} (m-n) \, \{2m^2+2n^2-mn-8\} \, L_{m+n}.   \label{W3wedgealgebra}
\end{eqnarray}

While algebra sl(3,R) $\oplus$ sl(3,R) is a generalization of diffeomorphism and local Lorentz symmetry  in Chern-Simons (CS) gauge theory,  this algebra sl(3,R) $\oplus$ sl(3,R)  is also  a symmetry algebra of boundary CFT as well as the bulk, where each sl(3,R) is a global W$_3$ algebra for rightmover and leftmover, respectively. 
In this paper we will address a new renormalization group flow in W$_3$ confromal field theory (CFT) and its description in terms of a geometry of  higher dimensional space for spin-3 gravity in AdS$_3$ space, which is  obtained by combining an `internal space' and an ordinary spacetime. In CFT operators are assigned to each point in the space by moving by $L_{-1}$ and its anti-holomorphic partner $\overline{L}_{-1}$. In W$_3$ extended CFT $W_{-2}$ and $W_{-1}$ also move operators in the `internal space'.\footnote{This space is associated with spin-3 gauge transformation. It is actually  not an internal space; its coordinates mix with those of the real space under W$_3$ transformations. However, in this paper we will call this space an `internal space'.}

In AdS$_{d+1}$ gravity the spacetime can be embedded as a hypersurface into a d+2 dimensional Minkowski  space $M^{d,2}$. 
This flat space has SO(d,2) symmetry and at any point on the hypersurface we have an invariant subgroup SO(d,1).  In the case of AdS$_3$, SO(2,2)=SL(2,R) $\otimes$ SL(2,R) and the symmetry at a point in the bulk is SL(2,R). By using this property a localized state of a scalar field put inside the bulk of AdS$_3$ space was constructed in \cite{Miyaji}\cite{V}\cite{NO}\cite{GT}. There have also been attempts to reconstruct bulk fields from the boundary CFT data by using bulk to boundary propagators.\cite{HKL}\cite{HKLL1}\cite{HKLL2} 
In the case of 3d spin-3 gravity, however,  we do not have embedding of the curved AdS spacetime including the internal space for W$_3$ gauge symmetry  within a higher-dimensional flat space. Best we can do is to make an assumption on the invariant subgroup according to symmetry considerations.  At least for the localized state for a scalar field such an extension is straightforward. 

In the case of AdS$_3$ gravity, where isometry is SL(2,R) $\otimes$ SL(2,R), the condition for the localized state of a scalar field $|\psi\rangle$ put in the center of the bulk is given  in terms of the generators of a  sub-algebra sl(2,R) and given  by \cite{Miyaji}\cite{V}\cite{NO}\cite{GT}
\begin{eqnarray}
&& (L_n-(-1)^n \, \overline{L}_{-n}) |\psi \rangle=0 \quad (n=-1,0,1). \label{LbL}
\end{eqnarray} 
In the case of spin-3 gravity the symmetry algebra sl(3,R) $\oplus$ sl(3,R)  is generated by $L_n$, $\overline{L}_n$ ($n=0, \pm 1$) and $W_n$, $\overline{W}_n$ $(n=0,\pm 1, \pm 2)$. For this theory conditions for a localized state of a scalar filed with W charge will be expressed in terms of generators of the  sl(3,R) sub-algebra which contains the above sl(2,R) sub-algebra. Then the  conditions for localized state will be (\ref{LbL}) and\footnote{The localized state for a scalar field with distinct values of chiral and anti-chiral W charges $\mu$ and  $\bar{\mu}$, the right hand side of the equation for $n=0$ needs to be replaced by a non-zero constant. }
\begin{eqnarray}
&&(W_n-(-1)^n \, \overline{W}_{-n})|\psi \rangle=0 \quad (n=-2,-1, \ldots, 2).  \label{WbW}
\end{eqnarray}

The first purpose of this paper is to construct a localized bulk state of a scalar field with a W$_3$ charge in 3d spin 3 gravity by solving (\ref{LbL}) and (\ref{WbW}) and study the equation of motion for the scalar field. We will work in Poincar\'e coordinates and denote sl(3,R) generators as $L^h_n$ and $W^h_n$. The generators without superscript $h$, $L_n$ and $W_n$, are generators in the elliptic basis for the global coordinates\cite{GT}. These are related by (\ref{LPg}) and (\ref{WPg}). 
To write down the equation of motion it is necessary to introduce coordinates $\alpha^{\pm}$, $\beta^{\pm}$ and $\gamma$ for the `internal space'.\footnote{These are sources for $W^h_{-2}$, $\overline{W}^h_{-2}$, $W^h_{-1}$, $\overline{W}^h_{-1}$, and $W^h_0+\overline{W}^h_0$. See (\ref{Plocalized}).}  This higher-dimensional space is an analog of superspace in supersymmetric theory and this will be called W space in this paper.  Structure of the W space obtained by combining 3d bulk spacetime and this `internal space' will be elucidated.  Metric of this W space (\ref{8dmetric}) is shown to be  a solution to higher-dimensional Einstein equation (\ref{Einstein8d}) with negative cosmological constant.   
The second purpose is to obtain an infinite dimensional representation in the bulk of the wedge mode algebra (\ref{W3wedgealgebra}) in terms of differential operators $\partial_{x^{\pm}}$, $\partial_{\alpha^{\pm}}$, $\partial_{\beta^{\pm}}$, $\partial_{\gamma}$ in addition to ordinary derivatives $\partial_{x^{\pm}}$, $\partial_y$. Here $y$ is a radial coordinate. With this representation it is possible to obtain two- and three-point functions of quasi-primary operators $\Phi(x^+,\alpha^{+}, \beta^{+},x^-,\alpha^{-}, \beta^{-})$ at $x^{\pm}=t\pm x$ on the boundary and at $(\alpha^{\pm}, \beta^{\pm})$ in the `internal space', which are actually descendants created from a Virasoro quasi-primary field $\Phi(x^+,x^-)$ by $W^h_{-1}, W^h_{-2}$, $\overline{W}^h_{-1}$, and $\overline{W}^h_{-2}$. Similar type of correlation functions are computed by Fateev and Ribault in \cite{FateevRibault} by using their variables $x$, $y$, and $w$ for a representation of sl(3,R) different from ours. Even if these variables may be related to our $x$, $\alpha$ and $\beta$ by suitable change of variables, the relation of the two sets of variables are not known.  Hence it is necessary to construct the representation of sl(3,R) algebra in terms of $(x,\alpha,\beta)$ from scratch. 

By using the above results it will be argued that the combined space of the bulk space and the internal space describes a renormalization group (RG)  flow along $\beta^{\pm}$ coordinates as well as along holographic $y$ coordinate from an AdS$_3$ vacuum with $W_3$ symmetry to another AdS$_3$ vacuum. These two vacua are located on 3D hypersurfaces at $\beta^{\pm}=0$ and $\beta^{\pm}=\infty$, respectively. The AdS radii of the two AdS$_3$ spaces are different by a factor 2. 
This flow is also  confirmed by study of correlation functions. Two- and three-point functions of  W-descendants, which are quasi-primary fields moved in the internal space by $W^h_{-1}$ and $W^h_{-2}$, are computed.\footnote{Some properties of three-point functions of W-descendats are discussed in \cite{BW}.} Two point function of quasi-primary chiral operators $\Phi_h(x^+,\alpha^{+}, \beta^{+})$
at $\alpha^{+}=\beta^{+}=0$ have the form
\begin{equation}
\langle 0|\Phi_h(x_1^+,0, 0) \Phi_h(x_2^+,0, 0)|0\rangle=\frac{1}{(x_{12}^+ )^{2h}}. \qquad (x_{12}^+ \equiv x^+_1-x_2^+)
\end{equation}
When the background for $\beta^{+}$ is introduced, the two point function becomes complicated, (\ref{twopoint}), and when a limit $\beta^{+} \rightarrow \infty$ is taken, we have
\begin{equation}
\lim_{\beta^{+} \rightarrow \infty}   \langle 0|(\beta^+)^{h} \, \Phi_h(x_1^+,0, \beta^+) (\beta^+)^{h}\Phi_h(x_2^+,0, \beta^+)|0\rangle \nonumber \\
=\frac{1}{(x_{12}^+)^{h}}.
\end{equation}
Hence the conformal weight changes as $h \rightarrow h/2$. 

Similar RG flow was reported in \cite{Kraus}\footnote{Two vacua of spin-3 gravity is also discussed in \cite{Castro}.}, where  a solution to the equation of motion in sl(3,R) $\oplus$ sl(3,R) CS theory of  3d spin 3 gravity, which represents RG flow between W$_3$ and $W_3^{(2)}$\cite{Bilal} vacua, was obtained and it was shown that stress tensor $T_{UV}$ in the UV, which has conformal weights $(h,\bar{h})=(2,0)$, flows to an operator with $(h',\bar{h'})=(0,4)$ in the IR. Because the pattern of the change of the conformal weight is similar to our case, field theory is also expected to flow to W$_3^{(2)}$ vacuum in the RG flow of the present paper. In sec.7 it is also shown that $L^h_{-1}$ in $\beta^+=0$ vacuum flows to $W^h_{-2}$ at $\beta^+ \rightarrow \infty$. 
$W^h_{-2}$ is a `new $L^h_{-1}$' for $W_3^{(2)}$ theory. 
However, properties of our flow and that in \cite{Kraus} are distinct: our flow is associated with a transformation of quasi-primary operators ${\cal O} \rightarrow e^{\beta^+ W^h_{-1}} {\cal O} e^{-\beta^+ W^h_{-1}}$.
It will be argued in sec.7 that this flow is driven by the following additional terms in the action.
\begin{equation}
\Delta S=\beta^+ (-iW^h_{-1})+\beta^- (-i\overline{W}^h_{-1}).
\end{equation}
The generators $W^h_{-1}$ and $\overline{W}^h_{-1}$ are anti-hermitian, and their c-number counterparts are pure imaginary. 

This paper is organized as follows. In sec.2 a boundary state in Poincar\'e coordinates is related to the primary state at the origin of Euclidean plane. In sec. 3 a localized state for a scalar field in the bulk is constructed for 3D spin 3 gravity. In sec. 4 an infinite dimensional representation of generators in the Poincar\'e coordinates, $L^h_n (n=-1,0,1)$ and $W^h_n (n=-2,-1,0,1,2)$, is constructed. In sec. 5 scalar equation of motion is derived and the metric tensor of the 8 dimensional space obtained by combining 3d spacetime and the `internal space' is obtained. This turns out  a specific solution to the Einstein equation with negative cosmological constant. It is pointed out that this space describes a RG flow from one W$_3$ vacuum to another. In sec. 6 two- and three-point functions of quasi-primary holomorphic operators $\Phi(x^+,\alpha^+,\beta^+)$ on the boundary and the internal space are calculated by solving differential equations which comes from W$_3$ symmetry of the vacuum. In sec. 7 it is shown that these correlation functions show a RG flow along the $\beta^+$ coordinate. It is also argued that this RG flow is triggered by adding extra terms to the action.  Sec. 8 is left for summary. Two appendices A, B contain some technical details for obtaining costraints on the boundary state and the localized state in the bulk.

\section{A Boundary State}
\hspace*{5mm}
In this section we will consider a localized state of a scalar field, which is extrapolated to the boundary. This state $|\psi\rangle_B$ 
is obtained from $|\psi \rangle$ by the following transformation\cite{GT}.
\begin{eqnarray}
|\psi \rangle_B &=& \lim_{\rho \rightarrow \infty}  g(\rho) \, |\psi \rangle,  \label{psiB}\\
g(\rho) &=& e^{-\rho\frac{1}{2} \,  (L_1-L_{-1}+\overline{L}_1-\overline{L}_{-1})} \label{g}
\end{eqnarray}
The state on the boundary satisfies the following conditions in stead of 
(\ref{LbL}) and (\ref{WbW}).
\begin{eqnarray}
&& \lim_{\rho \rightarrow \infty} \, g(\rho) \, (L_n-(-1)^n \, \overline{L}_{-n}) \, g(\rho)^{-1}|\psi \rangle_B=0, \quad (n=-1,0,1), \label{LbLB} \\
&&\lim_{\rho \rightarrow \infty} \, g(\rho) \, (W_n-(-1)^n \, \overline{W}_{-n}) \, g(\rho)^{-1}|\psi \rangle_B=0 \quad (n=-2,-1,0,1,2)
\end{eqnarray} 
Let us first consider (\ref{LbLB}). 
By working out several commutators the equations in  (\ref{LbLB}) before the limit $\rho \rightarrow \infty$  are given by 
\begin{eqnarray}
&&\Big[(L_0-\overline{L}_0)\cosh \rho-\frac{1}{2} \ (L_1+L_{-1}-\overline{L}_1-\overline{L}_{-1}) \, \sinh \rho\Big] \ |\psi \rangle_B=0, \\
&&\Big[ \frac{1}{2} (L_1-L_{-1}+\overline{L}_{-1}-\overline{L}_1)+\frac{1}{2} (L_1+L_{-1}+\overline{L}_{-1}+\overline{L}_1)\cosh \rho \nonumber \\
&& \qquad \qquad \qquad - (L_0+\overline{L}_0) \sinh \rho\Big] \ |\psi \rangle_B=0, \\
&&\Big[ \frac{1}{2} (\overline{L}_1-\overline{L}_{-1}+L_{-1}-L_1)+\frac{1}{2} (L_1+L_{-1}+\overline{L}_{-1}+\overline{L}_1)\cosh \rho \nonumber \\
&& \qquad \qquad \qquad - (L_0+\overline{L}_0) \sinh \rho\Big] \ |\psi \rangle_B=0.
\end{eqnarray}
In the limit $\rho \rightarrow \infty$ the above three conditions degenerate into two independent ones, and after taking  appropriate  linear combinations these are given by
\begin{eqnarray}
&&\Big[ L_0-\frac{1}{2}(L_1+L_{-1})\Big] \,|\psi \rangle_B=0, \label{chiral1}\\
&&\Big[ \overline{L}_0-\frac{1}{2}(\overline{L}_1+\overline{L}_{-1})\Big] \ |\psi \rangle_B=0. \label{achiral1}
\end{eqnarray}
A solution to these conditions are constructed on the primary state
\begin{equation}
|O_{\Delta}\rangle = \lim_{t \rightarrow i\infty} O_{\Delta}(\phi=0,t )|0\rangle,  \label{PrimaryEuc}
\end{equation}
where $O_{\Delta}(\phi,t)$ is a scalar primary operator with a conformal weight $(h,\bar{h})=(\Delta/2, \Delta/2)$ and a scaling dimension $\Delta$ on the boundary. $\phi$ is an angular variable in the global coordinates. 
This state corresponds to a scalar operator put on the infinite past of the Euclidean boundary, and $|0\rangle$ is the vacuum in 
the global patch. This state satisfies\footnote{For a while we will consider only the holomorphic and anti-holomorphic global Virasoro algebras.}  
\begin{eqnarray}
L_0|O_{\Delta}\rangle &=&\frac{\Delta}{2} \ |O_{\Delta}\rangle, \qquad L_1|O_{\Delta}\rangle =0, \nonumber \\
\overline{L}_0|O_{\Delta}\rangle &=&\frac{\Delta}{2} \ |O_{\Delta}\rangle, \qquad \overline{L}_1|O_{\Delta}\rangle =0.
\end{eqnarray}

It turns out the boundary state $|\psi\rangle_B$ is given by
\begin{equation}
|\psi\rangle_B = e^{L_{-1}+\overline{L}_{-1}} |O_{\Delta}\rangle.   \label{Primary in Poincare}
\end{equation}
This is shown by directly checking  (\ref{chiral1})-(\ref{achiral1}). 
It can also be shown that the operator which acts on $|\psi \rangle_B$ in (\ref{chiral1}) agrees with the Virasoro generator $L^h_{1}$ in the hyperbolic representation for the Poincar\'e coordinates.\cite{GT}
\begin{eqnarray}
L^h_1 &=& -L_0+\frac{1}{2}(L_1+L_{-1}), \nonumber \\
L^h_0 &=& \frac{1}{2} \, (L_1-L_{-1}), \nonumber \\
L^h_{-1} &=&  L_0+\frac{1}{2}(L_1+L_{-1}) \label{LPg}
\end{eqnarray}
There are similar relations for $\overline{L}^h_n$. $L^h_0$ is anti-hermitian and $L^h_{\pm 1}$ are hermitian.
The state  $|\psi\rangle_B$ satisfies the quasi-primary conditions for Virasoro algebra in the Poincar\'e coordinates.
\begin{eqnarray}
&& L^h_1 \, |\psi\rangle_B =\overline{L}^h_{1} \, |\psi\rangle_B=0, \label{Pprimary1} \\
&& L^h_0 \, |\psi\rangle_B =  \frac{\Delta}{2} \,  |\psi\rangle_B, \qquad 
\overline{L}^h_{0} \, |\psi\rangle_B= \frac{\Delta}{2}  \, |\psi\rangle_B \label{Pprimary2}
\end{eqnarray}
It can be shown that the exponential operator which appears in (\ref{Primary in Poincare}) carries a quasi-primary state $|O_{\Delta}\rangle  $ at the origin of Euclidean plane in the global coordinates to the point $(t,\phi)=(0,0)$ in the Lorentzian Poincar\'e coordinates. 

Next we will consider the conditions (\ref{WbW})  related to W generators. The transformed operators for finite $\rho$ are presented in appendix A. By sending the conditions (\ref{WbW}) to the boundary, the following conditions for $|\psi\rangle_B$ are obtained in addition to (\ref{chiral1}) and (\ref{achiral1}).
\begin{eqnarray}
&&(W_2-4W_1+6W_0-4W_{-1}+W_{-2}) \, |\psi \rangle_B=0, \\
&&(\overline{W}_2-4\overline{W}_1+6\overline{W}_0-4\overline{W}_{-1}+\overline{W}_{-2}) \, |\psi \rangle_B=0.
\end{eqnarray}
As in the case of the global Virasoro generators, these conditions define $W^h_2$ and $\overline{W}^h_2$ for the Poincar\'e coordinates in terms of those for global coordinates. Then  the algebra (\ref{W3wedgealgebra})  determines remaining $W^h_n$'s.
\begin{eqnarray}
W^h_2 &=& \frac{1}{4} \, (W_2-4W_1+6W_0-4W_{-1}+W_{-2}), \nonumber \\
W^h_1 &=& \frac{1}{4} \, (W_2-2W_1+2W_{-1}-W_{-2}), \nonumber \\
W^h_0 &=& \frac{1}{4} \, (W_2-2W_0+W_{-2}), \nonumber \\
W^h_{-1} &=& \frac{1}{4}(W_2+2W_1-2W_{-1}-W_{-2}), \nonumber \\
W^h_{-2} &=& \frac{1}{4}( W_2+4W_1+6W_0+4W_{-1}+W_{-2}) \label{WPg}
\end{eqnarray}
Note that $W^h_n$ with even $n$ are hermitian operators and those with odd $n$ are anti-hermitian. 
A quasi-primary state of W$_3$ algebra at the origin of Euclidean plane will be denoted as 
\begin{equation}
|\Delta, \mu\rangle  = \lim_{t \rightarrow i\infty} O_{\Delta,\mu}(\phi=0,t)|0\rangle. 
\end{equation}
This state satisfies the primary state conditions
\begin{eqnarray}
&&L_1|\Delta, \mu\rangle=0, \nonumber \\
&& L_0|\Delta, \mu\rangle=\frac{\Delta}{2}|\Delta, \mu\rangle, \nonumber \\
&& W_2|\Delta, \mu\rangle=W_1|\Delta, \mu\rangle=0, \nonumber \\
&&W_0 |\Delta, \mu\rangle=\mu |\Delta, \mu\rangle \label{pm}
\end{eqnarray}
and similar conditions for $\overline{W}_n$.  The boundary state 
\begin{equation}
|\psi_{\Delta,\mu}\rangle_B= e^{L_{-1}+\overline{L}_{-1}}|\Delta,\mu\rangle \label{Wboundarystate}
\end{equation}
satisfies (\ref{Pprimary1})-(\ref{Pprimary2}) and 
\begin{eqnarray}
&& W^h_2|\psi_{\Delta,\mu}\rangle_B=W^h_1|\psi_{\Delta,\mu}\rangle_B=0, \nonumber \\
&&W^h_0 |\psi_{\Delta,\mu}\rangle_B=\mu |\psi_{\Delta,\mu}\rangle_B
\end{eqnarray}
and similar conditions with respect to $\overline{W}^h_n$. 

\section{Construction of A Localized State}
\hspace*{5mm}
We will solve the conditions (\ref{LbL}) and (\ref{WbW}) for the localized state $|\psi\rangle$. This state will be built on the boundary state $|\psi\rangle_B$. 

First we will consider the  case of AdS$_3$ and solve (\ref{LbL}) without W extension.  Although this state was solved in \cite{Miyaji} by a series expansion, we will  obtain the state here in terms of  an integral representation. The localized state is represented in a form
\begin{equation}
|\psi\rangle = \int dx d\bar{x} \, f(x,\bar{x}) \, 
e^{xL^h_{-1}+\bar{x}\overline{L}^h_{-1}} \, |\psi_{\Delta}\rangle_B.
\end{equation}
Let us first consider the condition (\ref{LbL}) with $n=0$. 
When $L_0$ is applied on $|\psi\rangle$, we have
\begin{eqnarray}
L^h_0 \, |\psi \rangle &=& \int dx d\bar{x} \, f(x,\bar{x}) \, L^h_0  \, 
e^{xL^h_{-1}+\bar{x}\overline{L}^h_{-1}} \, |\psi_{\Delta}\rangle_B \nonumber \\
&=& \int dx d\bar{x} \, f(x,\bar{x}) \, e^{xL^h_{-1}} \, 
\Big(e^{-xL^h_{-1}}L^h_0  \, e^{xL^h_{-1}}\Big) \, 
e^{\bar{x}\overline{L}^h_{-1}} \, |\psi_{\Delta}\rangle_B. 
\end{eqnarray}
We  work out the commutator $e^{-xL^h_{-1}}L^h_0  \, e^{xL^h_{-1}}=L^h_0+xL^h_{-1}$
and replace $L^h_0$ by $\Delta/2$ according to (\ref{Pprimary2}), and $L^h_{-1}$ by $\partial_x$ due to exponential $e^{xL^h_{-1}}$ in the integrand. It is assumed that the surface term drops out. 
Finally by performing partial integration, we have
\begin{eqnarray}
L^h_0 \, |\psi \rangle &=& \int dx d\bar{x} \, \Big(\frac{\Delta}{2}f(x,\bar{x}) -x\partial_x \,f(x,\bar{x})\Big) \, e^{xL^h_{-1}+\bar{x}\overline{L}^h_{-1}} \, |\psi_{\Delta}\rangle_B
\end{eqnarray}
By a similar rewriting of $\overline{L}^h_0 \, |\psi \rangle$,  the condition $(L^h_0-\overline{L}^h_0)|\psi\rangle=0$ yields an equation.
\begin{equation}
(x\partial_x-\bar{x}\partial_{\bar{x}}) \, f(x,\bar{x})=0  \label{3.4}
\end{equation}
The other two conditions in (\ref{LbL}) produce the following equations.
\begin{eqnarray}
&&(x^2\partial_x+\Delta x)\, f(x,\bar{x})=0, \\
&&(\bar{x}^2\partial_{\bar{x}}+\Delta \bar{x})\, f(x,\bar{x})=0
\end{eqnarray}
Solution to the above equations is given up to a multiplicative constant by
$f(x,\bar{x}) = (1+x\bar{x})^{\Delta-2}$, and the localized state is obtained. 
\begin{equation}
|\psi \rangle = \int dxd\bar{x} \, (1+x\bar{x})^{\Delta-2} \,  e^{xL^h_{-1}+\bar{x}\overline{L}^h_{-1}} \, |\psi_{\Delta}\rangle_B
\end{equation}
To carry out the integration, we put $x=re^{i\theta}$ and $\bar{x}=r e^{-i\theta}$. 
After expanding the exponential in powers of $r$ we carry out the $\theta$ integral. Then  $r$ integral is performed term by term via analytical continuation in $\Delta$, assuming that the sum  and $r$ integration can be exchanged.
\begin{equation}
|\psi \rangle =\frac{\pi}{1-\Delta} \, \sum_{n=0}^{\infty} \, \frac{(-1)^n}{n!\, (\Delta)_n} \, (L^h_{-1}\overline{L}^h_{-1})^n \, |\psi_{\Delta}\rangle_B
\end{equation}
Here $(\Delta)_n=\Delta (\Delta+1) \dots (\Delta+n-1)$. This expression of the localized state coincides with that obtained by a different method in \cite{Miyaji}. Equivalence with the HKLL construction is shown in Appendix A of \cite{GT}.

In the case of spin 3 gravity, the localized state is obtained in a similar way. 
It is obtained as an integral in a form
\begin{eqnarray}
|\psi\rangle &=& \int dxd\bar{x}dyd\bar{y}dzd\bar{z} \ F(x,y,z,\bar{x},\bar{y},\bar{z}) \, e^{xW^h_{-2}}e^{yW^h_{-1}}e^{zL^h_{-1}} \nonumber \\ && \qquad \qquad e^{\bar{x}\overline{W}^h_{-2}}e^{\bar{y}\overline{W}^h_{-1}}e^{\bar{z}\overline{L}^h_{-1}}\, |\psi_{\Delta,\mu}\rangle_B.  \label{W3localizedstate}
\end{eqnarray}
Here $\bar{x}$, $\bar{y}$, $\bar{z}$ are complex conjugates of $x$, $y$, $z$, respectively. These variables have nothing to do with the coordinates of spacetime.
They are used only in this section and appendix B. 
The function $F$ is obtained by the same method as above for AdS$_3$ gravity and the details are given in appendix B. The result is
\begin{equation}
F(x,y,z,\bar{x},\bar{y},\bar{z}) =\Big[\frac{R_+}{R_-}\Big]^{\frac{3}{4}\mu} \ T^{\frac{\Delta-8}{4}}
\label{localstateW}
\end{equation}
where 
\begin{eqnarray}
R_+ &=& 1+2\zeta\,  |1+\xi_2|^2 +\zeta^2\, |1-4\xi_1-2\xi_2-\xi_2^2|^2, \\
R_- &=& 1+2\zeta\,  |1-\xi_2|^2 +\zeta^2\, |1+4\xi_1+2\xi_2-\xi_2^2|^2, \\
T &=&1+4\zeta (1+|\xi_2|^2)
 +2\zeta^2 \, \Big\{|4\xi_1+2\bar{\xi_2}|^2+3|\xi_2^2-1|^2\Big\} \nonumber \\ && 
+4\zeta^3 \, \big(1+|\xi_2|^2\big)^{-1}
\Big\{ \big|1+|\xi_2|^4-\xi_2^2-\bar{\xi_2}^2\big|^2  \nonumber \\
&& \qquad \quad \qquad +\big|4(1+|\xi_2|^2)\xi_1+3\xi_2-\xi_2^3+\bar{\xi_2} 
+\xi_2^2\bar{\xi_2}\big|^2\Big\} \nonumber \\ &&
+\zeta^4\Big|1-16\xi_1^2-16\xi_1\xi_2-6\xi_2^2\xi_2^4  \Big|^2.
\end{eqnarray}
Here $\zeta$, $\xi_i$ are variables defined by
\begin{equation}
 \zeta=z\bar{z}=|z|^2, \qquad \xi_1=\frac{x}{z^2}, \qquad \xi_2= \frac{y}{z}, \qquad 
 \bar{\xi_1}= \frac{\bar{x}}{\bar{z}^2}, \qquad \bar{\xi_2}= \frac{\bar{y}}{\bar{z}}.
\end{equation}
Hence an integral representation of the localized state is obtained. 
$R_+$, $R_-$ and $T$ are manifestly positive definite. 
The integral in (\ref{W3localizedstate}) may be carried out by analytical continuation with respect to $\Delta$, and the solution to the condition (\ref{WbW}) will exist. The integral is not yet evaluated explicitly. 
Finally, notice that no relation between $\Delta$ and $\mu$ is required for existence of the localized state.

\section{Infinite Dimensional Representation of $L^h_n$ and $W^h_n$ in the Bulk}
\hspace*{5mm}
The localized state obtained in the previous section can be moved to arbitrary positions at $(x^+,x^-,y)$ in the bulk by unitary transformations. 
It is also possible to carry out extra W$_3$ transformations on this state. This will move the state  in the `internal space'. 
In the following we will work out this construction in Poincar\' e coordinates. 
As in \cite{GT} the explicit form of the localized state in the product of the bulk space and `internal space' will be given by
\begin{multline}
|\Phi (x^+, \alpha^+,\beta^+,x^-,\alpha^-,\beta^-,y,\gamma)\rangle \\
=e^{ix^+ L^h_{-1}} \, e^{ix^-\overline{L}^h_{-1}} \, e^{i\alpha^+W^h_{-2}} \, e^{i\alpha^-\overline{W}^h_{-2}} \, e^{\beta^+W^h_{-1}} \, e^{\beta^-\overline{W}^h_{-1}} \, e^{-\frac{i}{2} \, \gamma (W^h_0+\overline{W}^h_0)}\, y^{L_0^h+\overline{L}_0^h}\, |\phi\rangle,  \label{Plocalized}
\end{multline}  
where 
$x^{\pm}=t \pm x$ are light-cone coordinates on the boundary in Poincar\'e coordinates, $y$ is the radial coordinate, and $\alpha^{\pm}$ and $\beta^{\pm}$, $\gamma$ are parameters of the global W$_3$ transformation in the $c \rightarrow \infty$ limit.

By studying $SL(3,R) \otimes SL(3,R)$ transformation of this localized state at arbitrary point we will be able to obtain representation of $L^h_n$ and $W^h_n$ {\em in the bulk} in terms of differential operators. This is the purpose of this section. First, let us apply $e^{-\epsilon L^h_{-1}}$ on $|\Phi\rangle$, where $\epsilon$ is an infinitesimal parameter. We have
\begin{eqnarray}
&& e^{-\epsilon L^h_{-1}} \, |\Phi (x^+, \alpha^+,\beta^+,x^-,\alpha^-,\beta^-,y,\gamma)\rangle
=|\Phi (x^++i\epsilon, \alpha^+,\beta^+,x^-,\alpha^-,\beta^-,y,\gamma)\rangle \nonumber \\
&& \equiv e^{\epsilon \, \hat{L}^h_{-1}} \, |\Phi (x^+, \alpha^+,\beta^+,x^-,\alpha^-,\beta^-,y,\gamma)\rangle
\end{eqnarray}
and this defines a differential operator $\hat{L}^h_{-1}=i\partial_{x^+}$.\footnote{The flip of sign in front of $\epsilon$ is necessary to insure correct algebra.} 
For the other generators the exponential of the generator is moved to the right by using the W$_3$ algebra until it reaches the localized operator $|\phi\rangle$ at the central point, and the conditions (\ref{LbL}) and (\ref{WbW}) are used and then the exponential is moved back to the left.
By this procedure we identify the following representation. 
 \begin{eqnarray}
\hat{L}_{-1}^h &=& i\partial_+, \nonumber \\
\hat{L}_0^h &=& -x^+\partial_+-2\alpha^+\partial_{\alpha^+}-\beta^+\partial_{\beta^+}-\frac{1}{2}y\partial_y, \nonumber \\
\hat{L}_1^h &=& -i[(x^+)^2-3(\beta^+)^2]\partial_+ -ix^+y\partial_y-3i\beta^+\partial_{\gamma}+i[2(\beta^+)^3-4x^+\alpha^+]\partial_{\alpha^+}
\nonumber \\
&&-i[2x^+\beta^++4\alpha^+]\partial_{\beta^+}-iy^2 \cos (2\gamma) \partial_--iy^2\cos (2\gamma)\beta^-\partial_{\alpha^-}-iy^2\sin (2\gamma)\partial_{\beta^-}, \nonumber \\
&&
\end{eqnarray}
\begin{eqnarray}
\hat{W}_{-2}^h &=& i\partial_{\alpha^+}, \nonumber \\
\hat{W}_{-1}^h &=& -x^+\partial_{\alpha^+}-\partial_{\beta^+}, \nonumber \\
\hat{W}_0^h &=& 2i\beta^+\partial_++i [-(x^+)^2+(\beta^+)^2]\partial_{\alpha^+}-2ix^+\partial_{\beta^+}-i\partial_{\gamma}, \nonumber \\
\hat{W}_1^h &=& 3x^+\partial_{\gamma}+[4\alpha^+-6x^+\beta^+] \partial_++[(x^+)^3-3x^+(\beta^+)^2]\partial_{\alpha^+}+[3(x^+)^2-(\beta^+)^2] \partial_{\beta^+} \nonumber \\
&& -\beta^+y\partial_y-y^2 \sin (2\gamma)\partial_--\beta^-y^2\sin(2\gamma) \partial_{\alpha^-}+y^2 \cos (2\gamma)\partial_{\beta^-}, \nonumber \\
\hat{W}_2^h &=& -i \big[3(\beta^+)^4-(x^+)^4-16(\alpha^+)^2+6(x^+)^2(\beta^+)^2 \big]\partial_{\alpha^+} \nonumber \\
&&+ i \big[y^4+4y^2\beta^-\beta^+\cos (2\gamma)-4y^2\beta^-x^+\sin (2\gamma) \big]\partial_{\alpha^-} \nonumber \\ 
&&+i \big[16\alpha^+\beta^+-4x^+(\beta^+)^2+4(x^+)^3\big]\partial_{\beta^+} \nonumber 
\\ && +iy^2\big[4\beta^+ \sin (2\gamma)+4x^+\cos (2\gamma)\big]\partial_{\beta^-}-i \big[4(\beta^+)^3+12\beta^+(x^+)^2-16x^+\alpha^+\big]\partial_+ \nonumber \\
&&+iy^2 \big[4\beta^+\cos (2\gamma)-4x^+\sin (2\gamma)\big]\partial_-+i\big[6(\beta^+)^2+6(x^+)^2\big]\partial_{\gamma} \nonumber \\
&& +i\big[8\alpha^+-4x^+\beta^+\big] \, y\partial_y \label{differentialop}
\end{eqnarray}
Here $\partial_{\pm}$ denotes $\partial_{x^{\pm}}$ for simplicity. 
The expressions for $\hat{\overline{L}}_n^h$ and $\hat{\overline{W}}_n^h$ are obtained by an exchange $+ \leftrightarrow -$.
These generators satisfy the wedge-mode algebra (\ref{W3wedgealgebra}). 
In the remaining part of this paper, the `hat'  for the differential operators will be omitted for simplicity of notation.

\section{Scalar Equation of Motion and W Geometry as RG Flow}
\hspace*{5mm}
The localized state obtained in the preceding section satisfy a differential equation, which also depends on differential operators $\partial_{\alpha^{\pm}}$, $\partial_{\beta^{\pm}}$, and $\partial_{\gamma}$. This is derived by using quadratic Casimir operator of SL(3,R) like SL(2,R) case\cite{NO}\cite{GT}. This is given by
\begin{eqnarray}
C_2(L,W)&=&(L^h_0)^2-\frac{1}{2} \, (L^h_1 L^h_{-1}+L^h_{-1}L^h_1)+\frac{1}{8} \, (W^h_2W^h_{-2}+W^h_{-2}W^h_2)\nonumber \\
&&-\frac{1}{2} \, (W^h_1W^h_{-1}+W^h_{-1}W^h_1)+\frac{3}{4}(W^h_0)^2.
\end{eqnarray}
By using 
\begin{equation}
\Big(C_2(L,W)+C_2(\overline{L},\overline{W})\Big) \, |O_{\Delta,\mu}\rangle=\frac{1}{2} \, \Big\{ \Delta^2-8\Delta+3\mu^2\Big\} \,  |O_{\Delta,\mu}\rangle, 
\end{equation}
and the representation (\ref{differentialop}), Klein-Gordon equation in  W space is derived. 
\begin{eqnarray}
&& \Big[ y^2\partial_y^2-7y\partial_y-3\partial_{\gamma}^2-4y^2\cos 2\gamma \, \partial_{x^+}\partial_{x^-}-4y^2\cos 2\gamma \, (\beta^-\partial_{x^+}\partial_{\alpha^-}+\beta^+\partial_{x^-}\partial_{\alpha^+}  ) \nonumber \\
&&-4y^2\sin 2\gamma (\partial_{x^+} \partial_{\beta^-}+\partial_{x^-} \partial_{\beta^+})
-4y^2\sin 2\gamma (\beta^+\partial_{\alpha^+}\partial_{\beta^-}+\beta^-\partial_{\alpha^-}\partial_{\beta^+})\nonumber \\
&& -(y^4+4\beta^+\beta^-y^2 \cos 2\gamma)\partial_{\alpha^+}\partial_{\alpha^-}+4y^2 \cos 2\gamma \partial_{\beta^+}\partial_{\beta^-}-m^2\ell_{AdS}^2\Big] \, |\Phi\rangle=0, \label{KG}
\end{eqnarray}
where $m$ is a mass of the scalar field, $\ell_{AdS}$ AdS length,  and $m^2\ell_{AdS}^2= \Delta^2-8\Delta +3\mu^2$, 
which gives
\begin{equation}
\Delta = 4+\sqrt{m^2\ell_{AdS}^2+16-3\mu^2}.
\end{equation}
Eq (\ref{KG}) is an equation of motion for a scalar field propagating in an 8d spacetime  with the metric field given by
\begin{eqnarray}
ds_8^2 &=& {^{(8)}g}_{MN} \ dx^Mdx^N \nonumber \\
&=&y^{-2} \, dy^2-y^{-4}(y^2 \cos 2\gamma+4\beta^+\beta^-) dx^+dx^--4y^{-4}d\alpha^+d\alpha^-         \nonumber \\
&&+y^{-2} \cos 2\gamma d\beta^+d\beta^-+4y^{-4} (\beta^+dx^+d\alpha^-+\beta^-dx^-d\alpha^+)    \nonumber \\
&& -y^{-2}\sin 2\gamma (dx^+d\beta^-+dx^-d\beta^+)-\frac{1}{3} \, d\gamma^2, \label{8dmetric}
\end{eqnarray}
where $x^M=(x^+,x^-,y,\alpha^+,\alpha^-,\beta^+,\beta^-,\gamma)$. 
The determinant of this metric is $^{(8)}g=\text{det}\ {^{(8)}g}_{MN}=\frac{1}{12} y^{-18}$. It can be shown that  this metric satisfies Einstein equation, 
\begin{equation}
^{(8)}R_{MN}-\frac{1}{2}\,  {^{(8)}g}_{MN} {^{(8)}R}=-\Lambda_8 \, ^{(8)}g_{MN}  \label{Einstein8d}
\end{equation}
 with a negative cosmological constant $\Lambda_8=-36$, where the cosmological constant of the original AdS$_3$ is 
$\Lambda_3=-1$ $(\ell_{\text{AdS}}=1$ in our units). The metric (\ref{8dmetric}) is, however,  not the one of AdS$_8$ space. It has signature $ (+,+,+,+,-,-,-,-)$ and  isometry of (\ref{8dmetric}) is only SL(3,R) $\otimes$ SL(3,R). This is an analog of superspace in supersymmetric theory, because commutators of $W_m$'s are given in terms of  $L_n$ (\ref{W3wedgealgebra}).

The original physical AdS$_3$ space is the hypersurface embedded at $\alpha^{\pm}=\beta^{\pm}=\gamma=0$ in this W space. 
Other hypersurfaces with constant non-vanishing $\alpha^{\pm}$, $\beta^{\pm}$ and $\gamma$ may  also be considered. An induced metric on this hypersurface is given by 
\begin{eqnarray}
ds_8^2\Big|_{\text{hypersurface}} &=& y^{-2} \, dy^2-y^{-4}(y^2 \cos 2\gamma+4\beta^+\beta^-) dx^+dx^-.
\label{inducedmetric}
\end{eqnarray}
Property of the hypersurface depends on the values of $\gamma$ and $\beta^{\pm}$. 
\begin{itemize}
\item When $\beta^+,\beta^-=0$ and $-\frac{\pi}{4} < \gamma < \frac{\pi}{4}$, the hypersurface corresponds to an ordinary AdS$_3$ vacuum with AdS length $\ell_{\text AdS}$. 
\item When $\beta^+\beta^- >0$ and $\gamma =\pm \frac{\pi}{4}$, $\beta^{\pm}$ can be absorbed by rescaling of $y$ and  the hypersurface is again AdS$_3$ but with other value of AdS length $\ell'_{\text{AdS}}=\frac{1}{2} \, \ell_{\text{AdS}}=\frac{1}{2}$, because $y^{-2}dy^2=\frac{1}{4} \, (y^2)^{-2}(d(y^2))^2$.
\item For $\gamma \neq \pm \frac{\pi}{4}$ and $\beta^+\beta^- >0$ the spacetime on the hypersurface is not AdS, but asymptotically AdS. 
This is a solution interpolating two vacua: one corresponding to UV CFT at $y=0$ with a AdS length $\ell_{\text{AdS}}'$, and the other to IR CFT at $y=\infty$ with $\ell_{\text{AdS}}$.
Hence conformal symmetry is broken in the boundary field theory for non-zero $\beta^+\beta^-$.  
When $\beta^{\pm} \rightarrow \infty$, conformal symmetry is recovered.
\end{itemize}
This observation suggests that the parameters $\beta^{\pm}$ also play the role of holographic renormalization group scales, which the radial variable $y$ plays in the conventional AdS/CFT correspondence:\footnote{Restricting the bulk fields to a hypersurface with fixed values of $\alpha$, $\beta$, $\gamma$ is equivalent to transforming all the fields by  $e^{i\alpha W^h_{-2}}e^{\beta W^h_{-1}} e^{\frac{i}{2}\gamma W^h_0}$ with these values as fixed backgrounds. This transformation may be equivalent to adding some perturbations to the action integral.} 
when the values of $\beta^{\pm}$ are sent to infinity, then the dual CFT may flow  to another one. \footnote{In \cite{Kraus} a solution to equation of motion for sl(3,R) $\otimes$ sl(3,R) CS gauge theory which interpolates between two vacua was presented. 
This solution agrees with the induced metric (\ref{inducedmetric}). 
This fact may be an evidence that  (\ref{W3localizedstate}) is  a correct localized state in 3D spin-3 gravity.
} Metric (\ref{8dmetric}) appears to describe a structure of renormalization group flows between W$_3$ vacua.  It is stressed that the physical spacetime is a 3D hypersurface with constant $\alpha^{\pm}$, $\beta^{\pm}$ and $\gamma$ embedded in the 8D space. As the values of these variables are changed, the spacetime as well as the dual CFT change just like they do as the radial variable $y$ changes. 

This view point is interesting and worth further investigation. In sec. 6 two- and three-point correlation functions of quasi-primary operators in W$_3$-extended CFT are calculated, and in sec. 7 it is shown that introduction of infinite-valued $\beta^{\pm}$ background to the quasi-primaries actually modifies the conformal weights of quasi-primary fields.

\section{Correlation Functions on the Boundary}
\hspace*{5mm}
In the preceding section representation (\ref{differentialop}) of W-algebra generators in the bulk  in terms of differential operators are obtained. In this section this result will be used to 
to study correlation functions of quasi-primary operators. In the case of global Virasoro algebra the descendants of quasi-primary states $|O_{\Delta}\rangle$ are 
given by $(L_{-1})^n|O_{\Delta}\rangle$ ($n=1,2,...$) and these states are put into one to one correspondence with local states at arbitrary boundary points by a map $|\psi(x^+)\rangle = e^{ix^+ L_{-1}} |O_{\Delta}\rangle$. Actually, this new state is a quasi-primary state with respect to new generators $L_{1}(x^+)$ and  $L_0(x^+)$ at point $x^+$.  Hence descendants created by the raising operators of the wedge algebra plays an important role in the representation theory. In the case of W$_3$ algebra (in the $c \rightarrow \infty$ limit), $W_{-2}$ and $W_{-1}$ are such raising operators and it will be natural to introduce an `internal space' with coordinates $\alpha^{\pm}$ and $\beta^{\pm}$ in accord with (\ref{Plocalized}). Descendants generated by $W_{-2}$ and $W_{-1}$ are organized into quasi-primary states with respect to new generators $L_n(x^+,\alpha^+,\beta^+)$ and $W_n(x^+,\alpha^+,\beta^+)$ at point $(\alpha^+,\beta^+)$ in the `internal space', as well as point $x^+$ on the boundary. Because the commutators of $W_n$'s are $L_m$, the coordinates of this `internal space' mix with those of the real 2d spacetime, $x^{+}$, under SL(3,R) transformation. 

We need representation of the holomorphic SL(3,R) generators on the boundary. This is done by replacement 
\begin{eqnarray}
&&y\partial_y \rightarrow \Delta=2h, \\
&&\partial_{\gamma} \rightarrow -i\mu.
\end{eqnarray} 
and a subsequent limit $y \rightarrow 0$ in (\ref{differentialop}). 
\begin{eqnarray}
L^h_{-1} &=& i\partial_+, \nonumber \\
L_0^h &=& -x^+\partial_+-2\alpha^+\partial_{\alpha^+}-\beta^+\partial_{\beta^+}-\frac{1}{2}\Delta, \nonumber \\
L_1^h &=& -i[(x^+)^2-3(\beta^+)^2]\partial_+-3\mu\beta^++i[2(\beta^+)^3-4x^+\alpha^+]\partial_{\alpha^+}
\nonumber \\
&&-i[2x^+\beta^++4\alpha^+]\partial_{\beta^+} -i\Delta x^+
\end{eqnarray}
\begin{eqnarray}
W_{-2}^h &=& i\partial_{\alpha^+}, \nonumber \\
W_{-1}^h &=& -x^+\partial_{\alpha^+}-\partial_{\beta^+}, \nonumber \\
W_0^h &=& 2i\beta^+\partial_++i [-(x^+)^2+(\beta^+)^2]\partial_{\alpha^+}-2ix^+\partial_{\beta^+}-\mu, \nonumber \\
W_1^h &=&[4\alpha^+-6x^+\beta^+] \partial_++[(x^+)^3-3x^+(\beta^+)^2]\partial_{\alpha^+}\nonumber \\
&&+[3(x^+)^2-(\beta^+)^2] \partial_{\beta^+}  -\Delta \beta^+- 3i\mu x^+, \nonumber \\
W_2^h &=& -i \big[3(\beta^+)^4-(x^+)^4-16(\alpha^+)^2+6(x^+)^2(\beta^+)^2 \big]\partial_{\alpha^+} \nonumber \\
&&+i \big[16\alpha^+\beta^+-4x^+(\beta^+)^2+4(x^+)^3\big]\partial_{\beta^+} \nonumber 
\\ &&-i \big[4(\beta^+)^3+12\beta^+(x^+)^2-16x^+\alpha^+\big]\partial_+ \nonumber \\
&&+6\mu \big[(\beta^+)^2+(x^+)^2\big]  +i\Delta \big[8\alpha^+-4x^+\beta^+\big]  \label{differentialopbondary}
\end{eqnarray}
These generators satisfy sl(3,R) algebra (\ref{W3wedgealgebra}). 
Similar representation is presented in sec 15.7.4 of \cite{FMS}. \footnote{See also eqs (2.25)-(2.32) of  \cite{FateevRibault}. } This representation of sl(3,R), however,  differs from (\ref{differentialopbondary}). Although this representation may be made coincide with (\ref{differentialopbondary}) by change of variables, in order to find out appropriate way of changing variables, if any,   knowledge of  (\ref{differentialopbondary}) itself is necessary. 
Because variables $\alpha^+$ and $\beta^+$ correspond to $W^h_{-2}$ and $W^h_{-1}$ descendants, 
 (\ref{differentialopbondary}) must be used for our purpose.  

Let $O_{\Delta_i,\mu_i}(x_i^+,\alpha_i^+,\beta_i^+)$ $(i=1,2)$ be two holomorphic quasi-primary operators at points $(x_i^+,\alpha_i^+,\beta_i^+) $ in the W space.   
Due to invariance of the vacuum two-point function of these operators,
\begin{equation}
G_{12}(x_1^+,\alpha_1^+,\beta_1^+; x_2^+,\alpha_2^+,\beta_2^+)=\langle 0| T O_{\Delta_1,\mu_1}(x_1^+,\alpha_1^+,\beta_1^+)\ O_{\Delta_2,\mu_2}(x_2^+,\alpha_2^+,\beta_2^+)            |0 \rangle
\end{equation}
should satisfy the following differential equations. $T$ is the time-ordering prescription.\footnote{Correlation functions in W$_3$ CFT are computed by Fateev and Ribault in \cite{FateevRibault}  by using the same methods and by using representation of sl(3,R) in terms of their variables $(x,y,w)$. Connection of these variables to our $(x^+,\alpha^+,\beta^+)$ is not known.        } 
\begin{eqnarray}
\Big( L^{h \, (1)}_n+L^{h \, (2)}_n\Big) \, G_{12}=0, \label{2L}\\
\Big( W^{h \, (1)}_n+W^{h \, (2)}_n\Big) \, G_{12}=0 \label{2W}
\end{eqnarray}
Here superscripts $(1)$ and $(2)$ on $L^{h}_n $ and $W^{h}_n$ refer to each operator $O_{\Delta_i,\mu_i}$. This two-point function is non-vanishing, if and only if $\Delta_1=\Delta_2$ and $\mu_1+\mu_2=0$ are satisfied.  Condition stemming from $L^h_{-1}$ restricts $G_{12}$ to depend on $x_1^+$ and $x^+_2$ only through  a combination $x_{12}^+=x_1^+-x_2^+$. Similarly, due to the condition from $W^h_{-2}$, $G_{12}$ is a function of $\alpha_{12}^+=\alpha_1^+-\alpha_2^+$. The remaining equations in (\ref{2L}) and (\ref{2W}) determine $G_{12}$ up to an overall constant. For a time order $t_1>t_2$ it is given by
\begin{equation}
G_{12} = (D_{12})^{-\frac{1}{4}\Delta_1+\frac{3}{4}\mu_1} \, (D_{12}^{\ast})^{-\frac{1}{4}\Delta_1-\frac{3}{4}\mu_1}. \label{twopoint}
\end{equation}
Here $D_{12}$ and $D_{12}^{\ast}$ are defined by
\begin{eqnarray}
D_{12} &=& (x^+_1-x^+_2)^2+(\beta^+_1-\beta^+_2)^2+2i(x^+_1-x^+_2)(\beta^+_1+\beta^+_2)-4i(\alpha^+_1-\alpha^+_2), \label{D12}\nonumber \\
D_{12}^{\ast} &=& (x^+_1-x^+_2)^2+(\beta^+_1-\beta^+_2)^2-2i(x^+_1-x^+_2)(\beta^+_1+\beta^+_2)+4i(\alpha^+_1-\alpha^+_2). \nonumber \\
&&
\end{eqnarray}
Two-point function for other time ordering is obtained by $t \rightarrow t-i\epsilon$ prescription. 
For completeness transformation property of $D_{12}$ under (\ref{differentialopbondary}) is given as follows.
\begin{eqnarray}
\Big( L^{h \, (1)}_{-1}+L^{h \, (2)}_{-1}\Big)  \, D_{12}&=& 0, \nonumber \\
 \Big( L^{h \, (1)}_{0}+L^{h \, (2)}_{0}\Big) \, D_{12}&=& -(2+\Delta) D_{12}\nonumber \\
 \Big( L^{h \, (1)}_{1}+L^{h \, (2)}_{1}\Big)  \, D_{12}&=& -i [2(x^+_1+x^+_2)+2i(\beta^+_1-\beta^+_2)+\Delta_1 \, (x^+_1+x^+_2)  \nonumber \\
&&\qquad \qquad -3i\mu_1(\beta_1^+-\beta_2^+)] D_{12}, \nonumber \\
 \Big( W^{h \, (1)}_{-2}+W^{h \, (2)}_{-2}\Big) \, D_{12}&=&0, \nonumber \\
 \Big( W^{h \, (1)}_{-1}+W^{h \, (2)}_{-1}\Big) \, D_{12} &=& 0, \nonumber \\
\Big( W^{h \, (1)}_{0}+W^{h \, (2)}_{0}\Big) \, D_{12} &=& 0, \nonumber \\
 \Big( W^{h \, (1)}_{1}+W^{h \, (2)}_{1}\Big) \, D_{12} &=& i[2(x^+_1-x^+_2)+2i(\beta^+_1+\beta^+_2)+i\Delta_1 (\beta^+_1+\beta^+_2)\nonumber \\
&& \qquad \qquad \qquad -3\mu_1(x^+_1-x^+_2)] D_{12}, \nonumber \\
 \Big( W^{h \, (1)}_{2}+W^{h \, (2)}_{2}\Big) \, D_{12} &=& 4\Big[4i(\alpha^+_1+\alpha^+_2)-(i\beta^+_1+x_1)^2+(i\beta^+_2-x_2^+)^2\nonumber \\
&&-2(\beta^+_1)^2+2(\beta^+_2)^2+2i\Delta_1(\alpha^+_1+\alpha^+_2) \nonumber \\ &&-i\Delta_1(x^+_1\beta^+_1+x^+_2\beta^+_2)\nonumber \\
&&+\frac{3}{2}\mu_1\{(\beta_1^+)^2-(\beta_2^+)^2+(x_1^+)^2-(x_2^+)^2\}\Big]D_{12}, 
\end{eqnarray}
where $\Delta_2=\Delta_1$ and $\mu_2=-\mu_1$ are used. 

Let us now consider three-point function $G_{123}$ of $O_{\Delta_i,\mu_i}(x_i^+,\alpha_i^+,\beta_i^+)$'s $(i=1,2,3)$, although analysis in this paper is restricted to $c \rightarrow \infty$ limit. This is interesting in W$_3$ CFT in its own right. 
\begin{eqnarray}
&&G_{123} \, (x_1^+,\alpha_1^+,\beta_1^+; x_2^+,\alpha_2^+,\beta_2^+;x_2^+,\alpha_2^+,\beta_2^+) \nonumber \\
&&=\langle 0| T\, O_{\Delta_1,\mu_1}(x_1^+,\alpha_1^+,\beta_1^+)\ O_{\Delta_2,\mu_2}(x_2^+,\alpha_2^+,\beta_2^+)\ O_{\Delta_3,\mu_3}(x_3^+,\alpha_3^+,\beta_3^+)            |0 \rangle
\end{eqnarray}
Like the two-point function, $G_{123}$ is a solution to the following  equations
\begin{equation}
\Big( \sum_{i=1}^3L^{h \, (i)}_n\Big) \, G_{123}=0, \qquad 
\Big( \sum_{i=1}^3W^{h \, (i)}_n\Big) \, G_{123}=0. \label{3LW}
\end{equation}
and $G_{123}$  is non-zero only for $\mu_1+\mu_2+\mu_3=0$.
Contrary to the case of global Virasoro algebra without W-algebra extension, there are three linearly independent solutions to these equations. 
\begin{eqnarray}
G_{123} &=& (D_{12}D_{12}^{\ast})^{-\frac{1}{8}(\Delta_1+\Delta_2-\Delta_3)} \, (D_{23}D_{23}^{\ast})^{-\frac{1}{8}(\Delta_2+\Delta_3-\Delta_1)} \nonumber \\
&& \qquad (D_{13}D_{13}^{\ast})^{-\frac{1}{8}(\Delta_1+\Delta_3-\Delta_2)} \, K,  \label{threepoint}
\end{eqnarray}
\begin{equation}
K = \left\{\begin{array}{c}
 \Big(\frac{D_{13}}{D_{13}^{\ast}}\Big)^{\frac{3}{4}\mu_1} \, \Big(\frac{D_{23}}{D_{23}^{\ast}}\Big)^{\frac{3}{4}\mu_2}, \ \text{or}\\
 \Big(\frac{D_{21}}{D_{21}^{\ast}}\Big)^{\frac{3}{4}\mu_2} \, \Big(\frac{D_{31}}{D_{31}^{\ast}}\Big)^{\frac{3}{4}\mu_3}, \ \text{or}\\
 \Big(\frac{D_{32}}{D_{32}^{\ast}}\Big)^{\frac{3}{4}\mu_3} \, \Big(\frac{D_{12}}{D_{12}^{\ast}}\Big)^{\frac{3}{4}\mu_1} 
\end{array}
\right.
\end{equation}
For $K$ linear combination of these terms is possible. There is, however, some restriction depending on the operators.
When one of the operator is an identity, a three-point function must reduce to a two-point function. If $O_{\Delta_3,\mu_3}$ is an identity, $\mu_3=0$ and $\Delta_1=\Delta_2$. Then the solution in the first line of $K$ cannot be present, because the coordinates $x_3^{\pm}$ do not cancel. 

\section{RG Flow }
\hspace*{5mm}
In the previous section correlation functions of holomorphic quasi-primary operators at point
$(x^+, \alpha^+,\beta^+)$ with conformal weight $h=\frac{\Delta}{2}$ and 
W$_3$ charge $\mu$ are computed. 
It is interesting whether there exists a point where 
the conformal weight changes, if the parameters $\beta^{\pm}$ are changed. We will study this problem by investigation of two (and three) point functions.

The two-point function (\ref{twopoint}) is written in terms  of a function $D_{12}$ (\ref{D12}) and its complex conjugate $D_{12}^{\ast}$.  
If two operators are put on a hypersurface with $\alpha^+=\alpha_0$ and $\beta^+=\beta_0$, then two operators have 
$\alpha_i^+=\alpha_0$ and $\beta_i^+=\beta_0$ $(i=1,2)$, and $\alpha_0$ disappears from $D_{12}$, while $\beta_0$ remains:
\begin{equation}
D_{12} \rightarrow  (x_1^+-x_2^+)^2+4i(x_1^+-x_2^+) \, \beta_0=(x_1^+-x_2^+) \, [x_1^+-x_2^++4i \beta_0] \label{Dbeta}
\end{equation}
For $\beta_0=0$ this corresponds to conformal two-point function with conformal dimension $\Delta_1$. 
However, the background  with non-vanishing $\beta_0$ breaks conformal invariance.\footnote{Note that here we fixed $\beta^+=\beta_0$ and do not treat it as a coordinate.} Conformal invariance is recovered by taking $\beta_0$ to infinity after changing the normalization of the quasi-primary field by multiplication of an appropriate power of $\beta_0$, {\em i.e.}, $\beta_0^{\frac{1}{2}\Delta_1}$. Because the factor which contains $\beta_0$ cancels with this multiplication factor in the two-point function (\ref{twopoint}), the conformal dimension changes from $\Delta_1$ to $\frac{1}{2}\Delta_1$ and W$_3$ charge $\mu$ vanishes at the end of the limit. 
Same conclusion can be reached for the three-point function (\ref{threepoint}).
Hence the CFT flows to another one. 

To summarize, operators  
\begin{equation}
\hat{\Phi}(x^+,\beta_0)=\beta_0^{\frac{1}{2}\Delta} \, e^{ix^+L^h_{-1}}e^{\beta_0W^h_{-1}} \, e^{L_{-1}}
 \Phi(0)e^{L_1}e^{-\beta_0W^h_{-1}}e^{-ix^+L^h_{-1}}, \label{hatPhi}
\end{equation}
where $\Phi(0)$'s are holomorphic primary operators with $h=\frac{\Delta}{2}$ at the origin of Euclidean plane, have conformal correlation functions of conformal weight $h'=\frac{1}{2}h$ in the limit $\beta_0 \rightarrow \infty$.
This RG flow proceeds along $\beta$ direction at $y=0$ in the higher dimensional space with metric (\ref{8dmetric}). This is similar to the holographic RG flow in the radial direction $y$. 
Because $\beta^{\pm}$ appears in the metric (\ref{8dmetric}) in the combination $\beta^{+}\beta^-/y^4$ with $y$, the limit $\beta^{\pm} \rightarrow \infty$ corresponds to $y \rightarrow 0$, {\em i.e.} UV. 

In \cite{Kraus} and \cite{Castro} two vacua  corresponding to two types of embedding of sl(2,R) algebra into sl(3,R) are studied. One embedding, a principal one,  employs $(L_1,L_0,L_{-1})$ as generators of sl(2,R) and corresponds to the standard $W_3$ algebra. In  the other embedding called a non-principal one,  $(\hat{L}_1,\hat{L}_0,\hat{L}_{-1})=  (\frac{1}{4}W_2,\frac{1}{2} L_0, -\frac{1}{4}W_{-2})$ generates sl(2,R) and this embedding gives rise to a new algebra known as $W^{(2)}_3$\cite{Bilal}.  On the gravity side an interpolating solution  connecting these two vacua was found in \cite{Kraus}.  In the non-principal embedding $W_{\pm 2}$ and $L_0$ correspond to the  stress tensor $T$, and  the other generators of sl(3,R) to a U(1) current $U$ and two spin $3/2$ currents $G_{\pm}$.  Hence the conformal weights of the currents $h$ changes to $h'=h/2$. Because  this pattern of the change of conformal weights is the same, this $W^{(2)}_3$ vacuum will also be the one to which  CFT will flow in our RG. Our  RG flow will, however,  be triggered by some additional  terms in the action, which is different from those in \cite{Kraus}. 

We will now argue that these additional terms to the action are given by
\begin{eqnarray}
\Delta S &=& -i\beta^+ W^h_{-1}-i\beta^- \overline{W}^h_{-1} \nonumber \\
&=& -\frac{i}{4}\beta^+ (W_2+2W_1-2W_{-1}-W_{-2})-
\frac{i}{4}\beta^- (\overline{W}_2+2\overline{W}_1-2\overline{W}_{-1}-\overline{W}_{-2}) \nonumber \\&&
\end{eqnarray}
There are $(-i)$'s, because $W^h_{-1}$  is anti-hermitian operator and its c-number counterpart is pure imaginary. 
In what follows only the holomorphic part is considered. 
Recall that under the flow an operator ${\cal O}(x)=e^{ix^+L^h_{-1}} {\cal O}(0) e^{-ix^+L^h_{-1}}$ transforms as\footnote{See (\ref{hatPhi}). Here we changed notation and 
${\cal O}(0)$ stands for an operator on the point $x^{\pm}=t \pm \phi=0$ of the cylindrical boundary.} 
\begin{eqnarray}
{\cal O}(x) &\rightarrow& e^{ix^+ L^h_{-1}} e^{\beta^+ W^h_{-1}} {\cal O}(0) e^{-\beta^+ W^h_{-1}} e^{-ix^+L^h_{-1}} \nonumber \\ 
&=& e^{\beta^+ W^h_{-1}} \, e^{ix^+\bm{L}^h_{-1}}{\cal O}(0)
e^{-ix^+ \bm{L}^{h}_{-1}}\, e^{-\beta^+ W^h_{-1}} \nonumber \\
&\equiv & e^{\beta^+ W^h_{-1}} \, \bm{{\cal O}}(x) \, e^{-\beta^+ W^h_{-1}}
\end{eqnarray}
Here $\bm{L}^h_n\equiv e^{-\beta^+ W^h_{-1}}L^h_n e^{\beta^+ W^h_{-1}}$. Notice that $W^h_{-1}=\bm{W}^h_{-1}\equiv e^{-\beta^+ W^h_{-1}}W^h_{-1} e^{\beta^+ W^h_{-1}}$. Because $L^h_{-1}$ and $W^h_{-1}$ do not commute, after the 
transformation by $e^{\beta^+W^h_{-1}}$ sl(3,R) generators change. The fact that $L^h_{-1}$ changes to $\bm{L}^h_{-1}=L^h_{-1}-\beta^+W^h_{-2}$
 is consistent with the expectation that when  additional terms are added to the action, they will give rise to  new contributions to the stress tensor. Furthermore for $\beta^+ \rightarrow \infty$, $\bm{L}^h_{-1}$ corresponds to the Virasoro generator for W$_3^{(2)}$ vacuum. $\bm{{\cal O}}(x)$ is an abbreviation for the operator shifted by $\bm{L}^h_{-1}$.  

Then arbitrary states of form $|\psi\rangle = {\cal O}_1(x^+_1) {\cal O}_2(x^+_2) \cdots {\cal O}_n(x^+_n) |0\rangle$ are transformed according to
\begin{equation}
|\psi \rangle \rightarrow e^{\beta^+ W^h_{-1}}\bm{ |\psi \rangle},
\end{equation}
where $\bm{|\psi\rangle} = \bm{{\cal O}}_1(x^+_1) \bm{{\cal O}}_2(x^+_2) \cdots \bm{{\cal O}}_n(x^+_n) |0\rangle$.
In the path-integral formalism this transformation  is achieved by adding to the action a term $\Delta S=-i\beta^+ W^h_{-1}$.  As  mentioned above, we regard  this additional term in the action  as the origin of the change of the translation generator $L^h_{-1} $. 

In the case of a similar transformation generated by $L^h_{-1}$ 
\begin{equation}
{\cal O}_i \rightarrow e^{i \delta x^+ L^h_{-1}} {\cal O}_i e^{-i \delta x^+ L^h_{-1}}
\end{equation}
states $|\psi\rangle$ are transformed as
\begin{equation}
|\psi \rangle \rightarrow e^{i \delta x^+L^h_{-1}} |\psi\rangle=e^{i \delta x^+ (H^h-P^h)}|\psi \rangle,
\end{equation}
where $H^h$ and $P^h$ are Hamiltonian and a momentum operator, respectively. These operators translate the system along the boundary directions, but do not alter the system. There is no flow.

In the flow discussed in this paper the stress tensor $T_{++}$, which is a quasi-primary operator with $h=2$, will flow to a current $U_+$, which has $h=1$. This can be easily shown by the same method mentioned above.  Then is there a stress tensor $T'_{++}$ at the end of flow? If the limit is also a CFT, it must exist, and this would flow from an $h=4$ current in the IR ({\em i.e.}, first UV).  In general there is no such quasi-primary operator in W$_3$ extended CFT.  Although there are $h=4$ operators $\Lambda$ and $\partial^2 T$,  states created by these are not quasi-primary: 
\begin{eqnarray}
L_1 |\Lambda(0)\rangle &=& 5|\partial T(0)\rangle ,  \qquad 
L_0 |\Lambda(0)\rangle =  4|\Lambda(0)\rangle,  \nonumber \\
W_2 |\Lambda(0) \rangle &=&  |T(0)\rangle,  \qquad 
W_1  |\Lambda(0) \rangle = |W(0)\rangle,  \qquad 
W_0  |\Lambda(0) \rangle = 3|W(0)\rangle \label{La}
\end{eqnarray}
\begin{eqnarray}
L_1 |\partial^2T(0)\rangle &=&|\partial T(0)\rangle ,  \qquad 
L_0 |\partial^2T(0)\rangle = 4|\partial^2 T(0)\rangle,  \nonumber \\
W_2 |\partial^2T(0) \rangle &=& 0, \qquad 
W_1  |\partial^2T(0) \rangle = |W(0) \rangle, \qquad 
W_0  |\partial^2T(0) \rangle = |\partial W(0)\rangle \label{pT}
\end{eqnarray}
Here $|T(0)\rangle \equiv \lim_{z \rightarrow 0}T(z)|0\rangle$.  $W(z)$ and $\Lambda(z)$ are the W current and the normal-ordered product of  $T$,  $\Lambda=(T,T)-(3/10)\partial^2 T$, respectively.
At present ,it is not clear with our method  if some linear combination of $\Lambda$ and $\partial^2 T$ would flow to the new stress tensor $T'$. This question may be answered by constructing two-point functions which are compatible with these  conditions (\ref{La}), (\ref{pT}) and taking $\beta \rightarrow \infty$ limit. Similarly, the state created by $W$ is not also quasi-primary ($W_1|W(0)\rangle =-20|T(0)\rangle$, {\it etc.}) and it is not known if two  $h=3/2$ currents could be obtained at the end of flow. This analysis is left for future study.

\section{Summary}
\hspace*{5mm}
In this paper 3D spin-3 gravity is studied from the point of view of bulk reconstruction and it is demonstrated explicitly  that the localized state can be constructed for spin-3 gravity (in an integral form (\ref{W3localizedstate}) with (\ref{localstateW})). For that purpose coordinates $\alpha^{\pm}$, $\beta^{\pm}$, $\gamma$ of the `internal space' which correspond to the wedge mode generators of W$_3$ algebra, $W_{-2}$, $W_{-1}$, $W_0$ are introduced. The equation of motion for a scalar field is written in W space including the `internal space' and  it is found that the metric of 8d gravitational background describes a RG flow between two AdS$_3$ vacua. This is confirmed by calculation of two- and three-point functions of Virasoro quasi-primary operators, which are W-descendants.
By taking $\beta^+ \rightarrow \infty$ limit, where $\beta^+$ is a source for $W_{-1}$ generator, the two-point functions of quasi-primary fields with conformal weight $h$  flow to those with conformal weight $h'=h/2$.  Further study of this RG flow is necessary.  It is interesting to investigate how do operators $W$ and $\Lambda$ flow and  whether it is  possible to start with $W^{(2)}_3$ vacuum and then flow back to $W_3$ vacuum. In this paper  it is found that by extending the spacetime to include the `$W$ directions', $L_n$ and $W_n$ transformations acquire a meaning of  diffeomorphism in the extended space in large $c$ limit, and representation of linear W$_3$ algebra in terms of differential operators is also constructed. This construction may be extended to the modes of $L$ and $W$ outside the wedge. To summarize, in the case of W$_3$ extended CFT, duality and holographic RG flow are realized in the extended higher-dimensional space. 

It is stressed that the physical spacetime is a 3D hypersurface embedded in the higher D spacetime with constant $\alpha$, $\beta$ and $\gamma$. Its metric is given by (\ref{inducedmetric}). When these variables are varied, then the bulk and  boundary theories change. When $\beta$ is changed, the boundary theory flows. $\alpha$ is a simple translation parameter. $\gamma$ can be absorbed into rescaling of $y$ as long as $\cos 2\gamma >0$. A bulk state of a scalar field on the 3D hypersurface with fixed $\beta$ is defined in terms of a boundary state on the same hypersurface through bulk reconstruction. 

Let us enumerate a few problems to study left for future. 
In this paper only the large $c$ limit is considered. How to incorporate 1/c corrections must be  studied. In this case modes outside the wedge such as $L_n \ (|n| \geq 2$) must be taken into account due to nonlinear terms in the algebra. 
In the gravity background studied in this paper spin-3 gauge field is not introduced. How to define spin-3 gauge field needs to be studied. 
Then interpretation of the spin-3 geometry as RG flows in the W space may be further elucidated. It is also interesting to find and understand W$_3$ black hole solutions in this space. 

\newpage
\setcounter{section}{0}
\renewcommand{\thesection}{\Alph{section}}
\section{Constraints on the Boundary State}
\hspace{5mm}
Transformation of $W_n$ in  (\ref{g}) is given by 
\begin{eqnarray}
g(\rho)\, W_2 \ g(\rho)^{-1} &=& \frac{1}{4}\big(\cosh^22\rho+2\cosh \rho+1\big)\, W_2-\sinh \rho \, \big(\cosh \rho+1\big) \, W_1  \nonumber \\
&&+\frac{3}{2}\sinh^22\rho W_0 
+W_{-1}\sinh \rho \, \big(1-\cosh \rho\big)\nonumber \\ &&+\frac{1}{4}\big(\cosh^22\rho-2\cosh \rho+1\big)W_{-2},
\end{eqnarray}
\begin{eqnarray}
g(\rho) \, W_1 \   g(\rho)^{-1} &=& \frac{-1}{8}\big(\sinh2\rho+2\sinh \rho\big)\, W_2+\frac{1}{2} \big(\cosh 2\rho+\cosh \rho\big) \, W_1-\frac{3}{4}\sinh2\rho W_0\nonumber \\ &&+\frac{1}{2}W_{-1}\big(\cosh2\rho-\cosh \rho\big)+\frac{1}{8}\big(-\sinh2\rho-2\sinh \rho\big)W_{-2},
\end{eqnarray}
\begin{eqnarray}
g(\rho)\, W_0 \  g(\rho)^{-1} &=& \frac{1}{8}\big(\cosh2\rho-1\big)\, W_2-\frac{1}{2} \sinh 2\rho \, W_1+\frac{1}{4}\big(3\cosh2\rho+1\big)W_0\nonumber \\ &&-\frac{1}{2}\sinh2\rho \, W_{-1}+\frac{1}{8}\big(\cosh2\rho-1\big) \, W_{-2},
\end{eqnarray}
\begin{eqnarray}
g(\rho) \,W_{-1} \ g(\rho)^{-1} &=& 
-\frac{1}{8}\big(\sinh2\rho+2\sinh \rho\big)W_{2}+\frac{1}{2}\big(\cosh2\rho-\cosh \rho\big)W_{1} -\frac{3}{4}\sinh2\rho W_0\nonumber \\ && +\frac{1}{2}W_{-1} \big(\cosh 2\rho+\cosh \rho\big)
-\frac{1}{8}\big(\sinh2\rho+2\sinh \rho\big)\, W_{-2},
\end{eqnarray}
\begin{eqnarray}
g(\rho) \,W_{-2} \ g(\rho)^{-1} &=& \frac{1}{4}\big(\cosh^22\rho-2\cosh \rho+1\big)W_{2}
+\sinh \rho\big(1-\cosh \rho\big)W_{1}+\frac{3}{2}\sinh^22\rho W_0 \nonumber \\ &&
+\frac{1}{4}\big(\cosh^22\rho+2\cosh \rho+1\big)\, W_{-2}- \sinh \rho \big(\cosh \rho+1\big)W_{-1},
\end{eqnarray}
Similar relations are obtained for barred (anti-holomorphic) generators. 

Then the boundary limits ($\rho \rightarrow \infty$)  of $W_n-(-1)^n\overline{W}_{-n}$ are given by
\begin{multline}
\lim_{\rho \rightarrow \infty}\ g(\rho) \, (W_2-\overline{W}_{-2}) \ g(\rho)^{-1}  \\ 
=\frac{1}{6} \, e^{2\rho}\ \Big(W_2-4W_1+6W_0-4W_{-1}+W_{-2} \
+4\overline{W}_{-1}-6\overline{W}_0 -\overline{W}_{-2}+4\overline{W}_1-\overline{W}_2\Big),
\end{multline}
\begin{multline}
\lim_{\rho \rightarrow \infty}\ g(\rho)\,  (W_1+\overline{W}_{-1}) \ g(\rho)^{-1}  \\ 
=\frac{1}{16} \, e^{2\rho}\ \Big(-W_2+4W_1-6W_0+4W_{-1}-W_{-2} \
+4\overline{W}_{-1}-6\overline{W}_0 -\overline{W}_{-2}+4\overline{W}_1-\overline{W}_2\Big),
\end{multline}
\begin{multline}
\lim_{\rho \rightarrow \infty} \ g(\rho)\,  (W_0-\overline{W}_{0}) \ g(\rho)^{-1}  \\ 
=\frac{1}{16} \, e^{2\rho} \ \Big(W_2-4W_1+6W_0-4W_{-1}+W_{-2} \
+4\overline{W}_{-1}-6\overline{W}_0 -\overline{W}_{-2}+4\overline{W}_1-\overline{W}_2\Big),
\end{multline}
By using these results the following two independent  conditions on the boundary state are obtained.
\begin{eqnarray}
&&(W_2-4W_1+6W_0-4W_{-1}+W_{-2}) \, |\psi \rangle_B=0, \\
&&(\overline{W}_2-4\overline{W}_1+6\overline{W}_0-4\overline{W}_{-1}+\overline{W}_{-2}) \, |\psi \rangle_B=0.
\end{eqnarray}

\section{Derivation of F (\ref{localstateW})}
\hspace{5mm}
In this section the function $R$ in (\ref{localstateW}) will be obtained. 
Condition $(L^h_0-\overline{L}^h_0)|\psi\rangle=0$ is transformed to an equation on $F$ by the same procedure as that used in deriving (\ref{3.4}):
\begin{equation}
(z\partial_z+y\partial_y+2x\partial_x)F=(\bar{z}\partial_{\bar{z}}+\bar{y}\partial_{\bar{y}}+2\bar{x}\partial_{\bar{x}})F. \label{L0}
\end{equation}
This equation implies that one of the six variables are redundant in the same way as  (\ref{3.4}) means $f(x,\bar{x})$ is a function only of $x\bar{x}$. 
Let us introduce the following five variables 
\begin{equation}
\zeta=z\bar{z}=|z|^2, \qquad \xi_1=\frac{x}{z^2}, \qquad \xi_2= \frac{y}{z}, \qquad 
 \bar{\xi_1}= \frac{\bar{x}}{\bar{z}^2}, \qquad \bar{\xi_2}= \frac{\bar{y}}{\bar{z}}. \label{newvariables}
\end{equation}
If  $F$ is regarded as a function of $\zeta$, $\xi_i$ and $\bar{\xi}_i$, then identities 
\begin{eqnarray}
(z\partial_z +y\partial_y+2x\partial_x)F&=& \zeta\partial_{\zeta} \, F, \nonumber \\
(\bar{z}\partial_{\bar{z}} +\bar{y}\partial_{\bar{y}}+2\bar{x}\partial_{\bar{x}})F&=& \zeta\partial_{\zeta} \, F 
\end{eqnarray}
hold. Hence (\ref{L0})  is automatically satisfied. 

Similarly, conditions from $L^h_1+\overline{L}^h_{-1}$, $W^h_0-\overline{W}^h_0$, $W^h_1+\overline{W}^h_{-1}$ and $W^h_2-\overline{W}^h_{-2}=0$    are given, respectively, by 
\begin{eqnarray}
&& \bullet \  [(y^3+3yz^2)\partial_x-(4x+6yz)\partial_y-(z^2+3y^2)\partial_z-\partial_{\bar{z}}+(\Delta-8)z \nonumber \\
&&\qquad \qquad \qquad +3\mu y+\bar{y}\partial_{\bar{x}}] F=0, \\
&&\bullet \ [(y^2+z^2)\partial_x-2z\partial_y-2y\partial_z-(\bar{y}^2+\bar{z}^2)\partial_{\bar{x}}+2\bar{z}\partial_{\bar{y}}+2\bar{y}\partial_{\bar{z}}]F= 0, \\
&& \bullet \ (4x-2yz)\partial_zF-(y^2+3z^2)\partial_yF+(2z^3-4xy)\partial_xF-\partial_{\bar{y}}F\nonumber \\
&& \qquad \qquad \qquad +[3\mu z+(\Delta-8) y]F=0, \\
&& \bullet \ [6\mu (z^2-y^2)-4\Delta yz-8\Delta x+32yz+64 x]F+[3z^4-6y^2z^2-y^4+16x^2]\partial_xF \nonumber \\
&&\qquad +[-4z^3+12y^2z+16xy]\partial_yF+[4yz^2+4y^3+16xz]\partial_zF+\partial_{\bar{x}}F=0.
\end{eqnarray}
and there are also complex conjugates of these equations. 
These equations are rewritten in terms of (\ref{newvariables}) as 
\begin{eqnarray}
&&\bullet \  -[1+3\zeta\xi_2^2+\zeta]\zeta\partial_{\zeta}F+\zeta[\xi_2^3+3\xi_2+2\xi_1+6\xi_1\xi_2^2]\partial_{\xi_1}F \nonumber \\
&&\qquad \qquad -\zeta[4\xi_1+5\xi_2-3\xi_2^3]\partial_{\xi_2}F+(2\bar{\xi}_1+\bar{\xi}_2)\partial_{\bar{\xi}_1}F+\bar{\xi}_2\partial_{\bar{\xi}_2}F \nonumber \\
&&\qquad \qquad  \qquad +(\Delta-8)\zeta F+3\mu \zeta\xi_2F=0, \\
&& \bullet  \ [1+4\xi_1\xi_2+\xi_2^2]\partial_{\xi_1}F-2[1-\xi_2^2]\partial_{\xi_2}F-2\zeta \xi_2\partial_{\zeta}F \nonumber \\
&& \qquad \qquad  -[1+4\bar{\xi}_1\bar{\xi}_2+\bar{\xi}_2^2]\partial_{\bar{\xi}_1}F+2[1-\bar{\xi}_2^2]\partial_{\bar{\xi}_2}F+2\zeta \bar{\xi}_2\partial_{\zeta}F=0, \\
&& \bullet \  [4\xi_1-2\xi_2]\zeta\partial_{\zeta}F+2[1-4\xi_1^2]F+[-3+\xi_2^2-4\xi_1\xi_2]\partial_{\xi_2}F-\zeta^{-1}\partial_{\bar{\xi}_2}F\nonumber \\
&&\qquad \qquad +[3\mu+(\Delta-8)\xi_2]F=0,\\
&& \bullet\ [3-2\xi_2^2-\xi_2^4+16\xi_1^2-8\xi_1\xi_2-8\xi_1\xi_2^3-32\xi_1^2]\partial_{\xi_1}F   + [-4+12\xi_2^2-4\xi_2^2-4\xi_2^4]\partial_{\xi_2}F \nonumber \\
&&\qquad \qquad +[4\zeta\xi_2+4\zeta\xi_2^3+16\zeta\xi_1]\partial_{\zeta}F+\zeta^{-2}\partial_{\bar{\xi}_1}F\nonumber \\ 
&&\qquad \qquad \qquad +[6\mu-6\mu \xi_2^2  +(32-4\Delta)(\xi_2+2\xi_1)]F=0.
\end{eqnarray}
Up to a multiplicative constant, solution to these equations and their complex conjugates are given by (\ref{localstateW}).

\end{document}